\documentclass[twocolumn,prl]{revtex4}

\usepackage[dvips]{graphicx}
\usepackage[dvips]{color}

\usepackage{graphicx}
\usepackage{epsfig}

\begin{document}
\title{Decoherence of Nuclear Spin Quantum Memory in Quantum Dot}
\author{Changxue Deng}
\thanks{Based on work presented at the 2004 IEEE NTC Quantum Device Technology Workshop.}
\author{Xuedong Hu}
\affiliation{Department of Physics, University at Buffalo, SUNY,
Buffalo, NY 14260-1500}

\begin{abstract}
  Recently an ensemble of nuclear spins in a quantum dot have been proposed as a long-lived 
quantum memory. A quantum state of an electron spin in the dot can be faithfully transfered into
nuclear spins through controlled hyperfine coupling.
Here we study the decoherence of this memory due to nuclear spin dipolar coupling and 
inhomogeneous hyperfine interaction 
during the {\it storage } period. We calculated the maximum fidelity of writing, 
storing and reading operations. Our results show that nuclear spin dynamics can severely 
limits the performance of the proposed device for quantum information processing and storage 
based on nuclear spins.
\end{abstract}
\maketitle

\section{introduction}

Electronic and nuclear spins in semiconductor nanostructures have been
studied intensively because of their potential applications in quantum 
computation and spintronics. In a series of proposals of solid state quantum computers,
electron or/and nuclear spins have been
suggested as qubits \cite{QC1,QC2}, the information carriers. Hyperfine coupling, the main interaction 
between an electron and the surrounding nuclei in semiconductors, can be utilized as a tool
to polarize the nuclei by optical 
pumping \cite{Gammon} or spin-polarized transport \cite{sb,dnp,Kawa}. 
On the other hand, it could also be a major decoherence channel for 
electron spins \cite{khaetskii,roger,saykin}.  

In a recent study, long-lived quantum memory of the electron spin in both
quantum dots \cite{taylor} and quantum Hall semiconductor nanostructures \cite{mani,mani-exp}, with the
help of hyperfine interaction, is proposed using highly polarized nuclei in these 
structures. The idea is to transfer the electron spin state to nuclear spins
coherently by tuning to the resonance condition of the nuclei-electron hyperfine flip-flop term;
this process is referred to as information writing. Since nuclear spins
have a long relaxation time $T_1$ (up to minutes), it is hoped that the information
of electron spin can be stored in nuclei for a  much longer period of time. The reverse
process (reading) can be activated on demand later by 
tuning back to the resonance point so as to transfer the collective nuclear spin state back to 
the electron spin.
Partially polarized nuclei which uniformly couple to the electron have
been shown not to depreciate the performance of the quantum memory dramatically \cite{taylor}.
Internal dynamics of nuclear spins has so-far been neglected.

However, nuclear spins are not static. Their evolution during the storage
time, when their flip-flop with the electron is turned off by either changing the
external magnetic field or simply removing the electron from the dot,
will induce decoherence in the quantum memory. In this paper we study the influence of nuclear 
spin dynamics on the information transfer and storage. Two types of spin interactions of 
nuclei could be in effect, namely the dipole-dipole coupling and the inhomogeneous 
hyperfine coupling with the electron (effective field seen by nuclei). 
During the writing and reading processes, the dipolar interaction can be ignored, 
since the operation time $\tau$ is determined by the hyperfine coupling constant 
(see below), so that $\tau H_d/\hbar$ is much less than 1, where $H_d$ is the dipolar 
Hamiltonian. However, it could be critical when the storage time approach the nuclear 
transverse relaxation time $T_2$, which is usually in the order of $10^{-3}$ to 
$10^{-4}$ second.   

In this paper we assume that initially the nuclear spins are fully polarized 
in a pure state $|I,\cdots,I \rangle$, where $I$ represent the magnitude of nuclear spins,
and $m_I$ means complete polarization. 
The system will undergo non-trivial evolution when a spin down electron is put into the 
dot. Once spin transfer is achieved at $\tau_w$, we turn off 
the resonance condition and let the nuclear spin system evolves for a long time $\tau_s$. 
Finally we read the information by putting back a spin up electron and turning on the
resonance. Our objective then is to study how the maximal reading fidelity vary as a 
function of the storage time $\tau_s$. 

\section{coherent spin dynamics of the effective two-level system}
In the writing and reading stage,
the electron-nuclei spin system can be approximated with a simple Hamiltonian
\begin{eqnarray}
H &=& \hbar \omega_e S_z - \hbar \omega_I \sum_{i=1}^{N} I_{iz} + H_h,
\nonumber \\
H_h &=& \sum_{i=1}^{N}A_i [I_{iz}S_z + \frac{1}{2}(I_{i+}S_- + I_{i-}S_+)].
\label{ham}
\end{eqnarray}
The first two terms in $H$ are the Zeeman energies of the electron spin and nuclear spins.
$H_h$ describes the hyperfine interaction in which a single electron interacts with an
ensemble of $N$ nuclear spins where $N\approx 10^4 - 10^5$ for a small quantum dot; 
$A_i = A|\psi(\mathbf{R}_i)|^2$ is the local strength of hyperfine coupling
and $\psi(\mathbf{R}_i)$ is the
electronic envelope wavefunction at the $i$th nucleus.
In the Hamiltonian $H$ we neglect the nuclear
dipole-dipole interaction as we have previously explained. The combined system of
nuclear and electron spins can be taken as an effective two-level system, where the 
electron spin could be either up or down. 
The time-dependent wavefunction reads
\begin{equation}
|\Psi(t) \rangle = \alpha(t) |\Psi_0 \rangle +
      \sum_{k=1}^{N} \beta_k(t) |\Psi_k \rangle,
\label{wave}
\end{equation}
where $|\Psi_0\rangle = |\Downarrow;I,\cdots,I\rangle$ and $|\Psi_k \rangle =
|\Uparrow;I,\cdots,(I-1)_k,\cdots,I\rangle$. Substituting Eq. \ref{wave} into
Sch\"{o}dinger equation, $i\hbar \partial \Psi (t)/\partial t = H \Psi (t)$,
we find
\begin{eqnarray}
i \dot{\alpha} &=& -(\frac{1}{2} \delta +\gamma) \alpha
+ \frac{\sqrt{2I}}{2}\sum_j \frac{A_j}{\hbar}\beta_j \nonumber \\
i \dot{\beta_k} &=& (\frac{1}{2} \delta -\gamma + \omega_I) \beta_k
- \frac{1}{2}\frac{A_k}{\hbar}\beta_k +
\frac{\sqrt{2I}}{2} \frac{A_k}{\hbar} \alpha,
\label{sch}
\end{eqnarray}
where $\delta=\omega_e + AI/\hbar$ and $\gamma=NI\omega_I$. Under the
resonance condition, the
electron-nuclei flip-flops are greatly enhanced since energies are conserved during these
processes. The resonance condition can be obtained by equating the total energies before
and after the scattering. We arrive at
\begin{equation}
\delta = \frac{A}{2N\hbar}-\omega_I.
\label{res}
\end{equation}
In deriving Eq. \ref{res} we have assumed that the hyperfine coupling constant is $A/N$.
Since $N\approx 10^5 \gg 1$ and $\omega_I \approx 10^{-3} \omega_e \ll \omega_e$, Eq. 
\ref{res} can be reduced to
$\delta =\omega_e+AI/\hbar \approx 0$, i.e., the resonance condition can be satisfied by adjusting the
external magnetic field.

An exact solution of Eq. \ref{sch} was found using Laplace transformation
\cite{khaetskii}. 
However, the non-trivial complex integral (the inverse Laplace transformation) in the solution
makes it less transparent. 
In the following we will apply a different approach. For later convenience we define
\begin{equation}
y_m = \sum_j \frac{A_j^m}{\hbar^m}  \beta_j.
\label{y}
\end{equation}
Using the resonance condition $\delta = 0$ and Eq. \ref{y}, Eq. \ref{sch} can be
transformed into
\begin{eqnarray}
i \dot{\alpha} &=& -\gamma \alpha + \frac{\sqrt{2I}}{2} y_1, \nonumber \\
i \dot{y}_m &=& -\gamma y_m - \frac{1}{2} y_{m+1} + \frac{\sqrt{2I}}{2}
\sum_j \frac{A_j^{m+1}}{\hbar^{m+1}} \alpha.
\label{wave1}
\end{eqnarray}
Without further approximation Eq. \ref{wave1} are equivalent to Eq. \ref{wave}. However,
if we write $y_2$ as
\begin{equation}
y_2 = \sum_j \frac{A_j^2}{\hbar^2} \beta_j \approx \frac{A}{N\hbar} y_1,
\label{y2}
\end{equation}
which is exact if the hyperfine coupling is uniform.
The group of differential equations are now closed with only $\alpha$ and $y_1$.
This approximation is appropriate because the decoherence induced
by the inhomogeneous hyperfine coupling is significant only when the time scale
is as long as $N/A$ for fully polarized nuclei \cite{khaetskii}. The spin transfer
time in which we are interested is much shorter. The argument will become
clear in the following calculations.

We consider two initial conditions corresponding to the writing and reading processes
respectively. For the initial state with a spin down
$\Psi(0) = |\Downarrow;I,\cdots,I\rangle$; $\alpha(0) = 1$ and
$\beta_1=\cdots=\beta_N=0$ and $y_1(0) = 0$. The solution of Eq. \ref{wave1} and Eq.
\ref{y2} is
\begin{equation}
\alpha(t) = e^{i\gamma t}\text{cos}(\frac{\Omega}{2} t),
\label{solution-alp}
\end{equation}
where $\Omega=\sqrt{2I\sum_j A_j^2}$ is the Rabi flopping frequency of the effective 
two-level system. Substituting Eq. \ref{solution-alp} back into Eq. \ref{sch} one find
the solution of $\beta_k$ is
\begin{equation}
\beta_k(t) = i \frac{A_k}{\sqrt{\sum_jA_j^2}} e^{i\gamma t}\text{sin}(\frac{\Omega}{2} t).
\label{solution-beta}
\end{equation}
We see that the system oscillates coherently between state  $| \Psi_0 \rangle$
and state $\sum_k A_k |\Psi_k \rangle/\sqrt{\sum_j A_j^2}$.
If we keep $y_2$ in Eq. \ref{wave1}, the solution will be different from Eq. \ref{solution-beta}.
However the corrections of the amplitude and the frequency for $\alpha$ are $\sim$ $1/\sqrt{N}$ and
$A/N$ respectively. In other words, the approximation Eq. \ref{y2} we made should be OK as long as 
the time scale is much smaller than $N/A$.
If we let $\tau_w = \pi/\Omega$, $|\alpha(t_w)|^2=0$ so that
complete spin transferring is achieved. The condition $\tau_w \ll N/A$ is
consistent with our approximation.
For the initial state with a spin up electron, $\Psi^{'}(0) \rangle =
\sum_k \beta^{'}_k(0) | \Psi_k \rangle$; $\alpha^{'}(0)=0$ and
$y_1(0) = \sum_j A_j \beta^{'}_j(0)/\hbar$. We find the solution in this case is
\begin{equation}
\alpha^{'}(t) = -i\hbar \frac{y_1(0)}{\sqrt{\sum_j A_j^2}} e^{i\gamma t}
\text{sin} (\frac{\Omega}{2} t).
\label{alp_prime}
\end{equation}
Suppose nuclear spin dynamics is frozen during the storage time,
$y_1(0)=\sum_j A_j\beta^{'}_j(0)/\hbar = \sum_j A_j\beta_j(\tau_w)/\hbar$ and
$|\alpha^{'}(\pi/\Omega)|^2=1$. However, if we allow the nuclear spins to evolve
during the storage time, $\beta^{'}_k(0)$ will be different from $\beta_k(t_w)$,
which will make $|\alpha^{'}(\pi/\Omega)|^2$ less than 1. In the next section,
we shall investigate how the decoherence induced by the nuclear spin dynamics
affect the coherent electron spin transfer in the read out process.

\section{nuclear spin dynamics}

During the period of storage we can either turn off the resonance by varying
the external magnetic field or simply remove the electron from the dot. In the first
case the electron-nuclei flip-flops are largely suppressed; nuclei only feel the
effective magnetic field generated by the term $A_iI_{iz}S_z$ which acts like
an inhomogeneous field for nuclei at different lattice sites. The effective
Hamiltonian of nuclear spins is
\begin{eqnarray}
H_n &=& \sum_i (\frac{A_i}{2} - \hbar\omega_I)I_{iz} + H_d, \nonumber \\
H_d &=& \sum_{i\ne j}2b_{ij}I_{iz}I_{jz} - \sum_{i\ne j} b_{ij}I_{i+}I_{j-}.
\label{Hn}
\end{eqnarray}
Here $H_d$ \cite{NMR} describe the secular terms of the dipole-dipole interaction;
$b_{ij}=\hbar^2\gamma_I^2(1-3\text{cos}^2\theta_{ij})R^{-3}_{ij}/4$, where 
$R_{ij}$ 
is
the distance between the nuclei at site $i$ and site $j$ and $\theta_{ij}$
is the angle between $\mathbf{R}_{ij}$ and the external magnetic field.
The second term in $H_d$ is the flip-flop term which introduces the nuclear 
spin diffusion \cite{NMR}. If the nucleus at the center of the dot is flipped down at 
$t=\tau_w$, then the down state of the nuclei will propagate to the nuclei 
at larger radius through their mutual flip-flops. After a sufficiently long period of time,
the nucleus with flipped spin state may be at the very boundary of the dot.
If one turns on the electron-nuclei resonance at this time, the injected 
spin-up electron has minimal chance to interact with the nucleus with flipped 
spin, because the probability of the electron at the boundary is negligible. 
This leads effectively to decoherence of the nuclear spin memory.

Since the truncated dipolar Hamiltonian $H_d$ conserve the total spin of all the
nuclei in the quantum dot, the general time-dependent spin wavefunction can
be expressed in terms of a linear combination of $|\psi_k \rangle$
\begin{equation}
|\psi(t)\rangle = \sum_k \beta_k(t) | \psi_k \rangle,
\end{equation}
where $|\psi_k \rangle = |I,\cdots,(I-1)_k,\cdots,I \rangle$.
The Sch\"{o}dinger equation is
\begin{equation}
i \dot{\beta}_{k} = \eta\beta_k -\frac{A_k}{2\hbar}\beta_k
-2I\sum_{m(k)}\frac{b_{km}}{\hbar}\beta_m,
\label{waven}
\end{equation}
where $\sum_{m(k)}$ means the summation over all $m$ except for $m= k$, 
\begin{eqnarray}
\eta &=&\frac{AI}{2} - (NI-1)\omega_I + 4I(I-1)\sum_{m(k)}\frac{b_{mk}}{\hbar} \nonumber \\
&+& 2I^2\sum_{m(k)}\sum_{n(m,k)}\frac{b_{mn}}{\hbar},
\label{eta}
\end{eqnarray}
which is a constant. The reason that we do not drop the second term on the right hand side
of Eq. \ref{waven}, although it is much less than $\eta$,
is that we are interested in the large time behavior ($\gg \pi/\Omega$) where non-uniform
hyperfine coupling may have significant effect. 

\begin{figure}[t]
\begin{center}
\epsfig{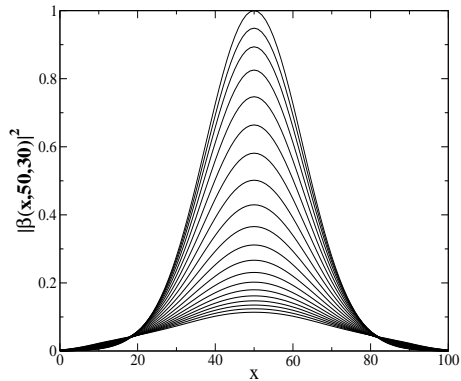}
\caption{$|\beta(x,50,30)|^2$ as a function of spatial coordinate $x$. The various curves 
in the figure represent the temporal profile of $|\beta|^2$ at different times starting 
from $t=0$. The time separation between every two neighboring curves is 1 millisecond.
It has been assumed that $L_r=25$ and $L_z=10$ in the simulations. The external magnetic
field is along the $x$ direction. The lattice
size which we use in these calculations is $101\times101\times61$. The coordinate system
has been established so that the center of the dot is located at (50,50,30).
 }
\label{bx}
\end{center}
\end{figure}
\begin{figure}[t]
\begin{center}
\epsfig{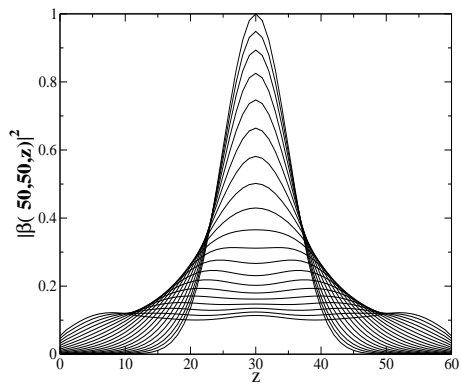}
\caption{
 $|\beta(50,50,z)|^2$ as a function of spatial coordinate $z$. All the parameters used 
in the calculations are the same as those in Fig. \ref{bx}.
 }
\label{bz}
\end{center}
\end{figure}

Even for such a simple situation, no analytical solution of Eq. \ref{waven} can be found. 
Therefore we have to rely on some numerical solutions. For this purpose we perform the following 
transformation for $\beta_k(t)$
\begin{equation}
\beta_k(t) = e^{-i(\eta-\frac{A_k}{2\hbar})t}\bar{\beta}_k(t).
\label{beta}
\end{equation}
Eq. \ref{waven} then becomes correspondingly
\begin{equation}
i \dot{\bar{\beta}}_k = -2I \sum_{m(k)} \frac{b_{km}}{\hbar}e^{i(A_m-A_k)t}\bar{\beta}_{m}.
\label{betabar}
\end{equation}
This set of transformed equations shows that the factor $\eta$ only contributes an overall 
phase factor for every $\beta_k$, so it will 
not affect our calculation of the probabilities. We set the
initial condition as $\bar{\beta}_k(0) = iA_k/\sum_jA_j^2$, i.e., the system starts right after the
spin of the electron is transfered to the nuclei (see. Eq. \ref{solution-beta}). We assume in 
our calculations that 
\begin{equation}
A(\mathbf{r}) = e^{-\frac{(x^2+y^2)}{L_r^2a^2}}e^{-\frac{z^2}{L_z^2a^2}},
\label{A}
\end{equation}  
where $a=5.65$ \AA ~is the lattice constant for GaAs. The coordinates $x,y$ and $z$ are in units of $a$.
$L_r$ and $L_z$ are two constants determined by the geometry of the quantum dot.
$I=3/2$ for all the nuclei in GaAs; $\gamma_{\text{As}}=4.58\times10^3\frac{1}
{\text{G}\cdot \text{s}}$. The time scale for dipolar interaction is $a^3/\gamma_I^2 \hbar \approx 10$
milliseconds. We use both
leapfrog and iterative Crank-Nicolson \cite{nm} method to solve Eq. \ref{betabar}.
We find that the leapfrog scheme is better in maintaining the unitarity of the coherent evolution for 
the nuclear spins, the total probability is preserved in the
evolution. We assume a closed boundary so that spin can not diffuse out of the dot. 
For simplicity we also assume that a single species of 
nuclei resides on a simple cubic lattice.

Let's now consider the case where the electron is removed from the dot so that the
evolution of the nuclear spins are only governed by the flip-flop term (nuclear spin 
diffusion). In Fig. \ref{bx} and Fig. \ref{bz} we show the probability
$|\beta|^2$ as a function of spatial coordinates $x$ and $z$.
The slightly different behaviors observed in Fig. \ref{bx} and Fig. \ref{bz} are due to
the sharper confinement along the $z$ direction; we have also integrate Eq. \ref{betabar} assuming
the external field is in the $z$ direction. Similar results as Fig. \ref{bx} and \ref{bz} are
found.
\begin{figure}[t]
\begin{center}
\epsfig{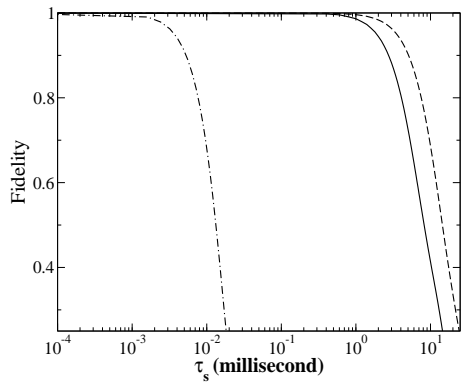}
\caption{Fidelity as a function of the storage time. The solid (dashed) represents the result 
of calculations when the external magnetic field is along $z$ ($x$) axis where there is no 
electron in the dot (no hyperfine coupling). The dot-dashed line shows the result when 
the hyperfine coupling is present. In all these calculations we assume a cubic lattice with size 
$101\times101\times61.$ 
 }
\label{fid}
\end{center}
\end{figure}
After time $\tau_s$, the coherent spin dynamics of electron can be activated by sending an electron
with up spin into the dot. 
To calculate the {\it fidelity} of the reading process, we use the definition given in Ref. \cite{taylor}
\begin{equation}
F = \text{Tr}\{ \rho (t_r) (\hat{S}_z- \frac{1}{2})^2\},
\end{equation}
where $\tau_r=\pi/\Omega$. The density operator is defined as $\rho(t) = |\Psi(t)\rangle \langle \Psi(t) |$, 
where the time-dependent wavefunction takes the form of Eq. \ref{alp_prime}. It is then found that
\begin{equation}
F = \hbar^2 \frac{|y_1(0)|^2}{\sum_j A_j^2}=\frac{|\sum_k\beta_k(t_s)A_k|^2}{\sum_j A_j^2}.
\label{F}
\end{equation}
The calculated fidelity is shown in Fig. \ref{fid} by the solid and dashed lines  
for different external magnetic field orientations. We see that the fidelity 
drops dramatically when $\tau_s$ reaches a few milliseconds. The difference of these two curves
can be understood as follows. Nuclear spin diffusion is stronger along the
external field direction. If this direction is the same as $z$ direction, which has larger
confinement (larger gradient) of the electronic wavefunction, the diffusion process is even faster. 
This explains that fidelity represented by the solid line ($z$) is less that that of the 
dashed line ($x$) at equal time $\tau_s$.

It appears therefore desirable to suppress the decoherence induced by spin diffusion with an inhomogeneous
hyperfine coupling \cite{roger,NSD}. This can be realized by leaving the electron in the dot during time $\tau_s$.
This suppression is due to energy non-conservation in the flip-flop
processes. However our result (the dot-dashed line in Fig. \ref{fid}) shows that the coherence
can only be maintained within a few microseconds, even worse than when the electron is removed. 
We can explain this result by identifying 
another decoherence channel, the inhomogeneous hyperfine coupling itself. The nucleus at different lattice
sites have different precession frequencies due to their interaction with the electron, so the phases
of the wavefunction $\beta_k(t)$ will not be uniform throughout the dot (see Eq. \ref{beta}). This 
induces destructive interference of the function $\sum_k A_k \beta_k(t)$. The time scale of
this decoherence is determined by the hyperfine coupling constant which is three order of magnitudes
larger than the dipolar constant.

\section{conclusion}
In conclusion, we have studied nuclear-spin dynamics induced decoherence of the long-lived quantum memory
proposed in Ref. \cite{taylor}.
We find that storage time is limited to a few milliseconds (without hyperfine coupling) and a few microseconds
(with hyperfine coupling $A_iI_{iz}S_z$). The performance of nuclear spin memory is limited by the two decoherence
channels: nuclear spin dipolar coupling and inhomogeneous hyperfine coupling. 
We only consider the fully polarized nuclei. Partially polarized nuclei give rise to
further decoherence. To make the device work as a truly {\it long-lived} quantum memory, one has to
improve the device by combining it with other techniques such as refocusing.

\end{document}